\documentclass[11pt]{article}
\usepackage[utf8]{inputenc}
\usepackage{amsmath}
\numberwithin{equation}{section}
\usepackage{amssymb}
\usepackage{hyperref}
\usepackage{graphicx}
\graphicspath{{./images/}}
\usepackage{float}
\usepackage{multirow}
\usepackage{booktabs}
\usepackage{bigstrut}
\usepackage{gensymb}
\usepackage{geometry}
\usepackage{fancyhdr}
\usepackage{physics}
\usepackage{bm}
\usepackage[dvipsnames]{xcolor}

\usepackage{subfig}

\def\p{\partial}
\def\f{\frac}

\def\mc{\mathcal}
\usepackage[T1]{fontenc}
\usepackage[utf8]{inputenc}
\usepackage{authblk}

\title{\Large{\bf Scalar fluid in scalar-tensor gravity from an equivalent picture of thermodynamic and fluid descriptions of gravitational dynamics}}
\author[1]{Abhinove Nagarajan Seenivasan\thanks{abhinove523@gmail.com, abhinovens@rnd.iitg.ac.in}}
\author[1]{Sayan Chakrabarti\thanks{sayan.chakrabarti@iitg.ac.in}}
\author[1]{Bibhas Ranjan Majhi\thanks{bibhas.majhi@iitg.ac.in}}
\affil[1]{\it Department of Physics, Indian Institute of Technology Guwahati, Guwahati 781039, Assam, India}

\begin{document}
  \maketitle

\begin{abstract}
We revisit the thermodynamic description of fluid, represented by scalar field in scalar-tensor gravity theory through a general approach to study the thermodynamics of relativistic fluids. In order to identify the fluid energy-momentum tensor, contrary to the existing way, we use equivalent description of gravitational dynamics both in Jordan and Einstein frames as thermodynamical identity and fluid equation on a generic null surface. Such an approach provides the energy-momentum tensors for scalar fluid as that of an ideal fluid in both the frames. We then mention few issues in the existing way of using Eckart's formalism for an ideal fluid. Our investigation suggests that the Eckart's frame may not be suitable to consider ideal fluid. Consequently a possible alternative description, valid in other than Eckart's frame, is being suggested for a general ideal fluid. Based on this we obtain a unified thermodynamic description in both Jordan and Einstein frames from our identified energy-momentum tensors. Finally, the relations between thermodynamic entities on different frames are being put forwarded. Thus, the description in either of the frames is enough to provide the complete picture.

\end{abstract}

%\end{titlepage}
\clearpage 
\fancyhead[L]{}
\fancyhead[R]{}

\parindent 0 ex

\tableofcontents

\fancyhead[L]{}
\fancyhead[R]{}

\newpage 

\section{Introduction and motivation}

Among the four fundamental forces of nature, gravity still remains the most difficult one to understand at the very fundamental level. The present understanding of gravity, as provided by Einstein's theory of General Relativity (GR) is based upon the description of spacetime geometry due to presence of matter. Irrespective of huge success of GR, both in the theoretical as well as experimental fronts,  a quantum theory of gravity still remains elusive. On another frontier, it was  found that \cite{Padmanabhan:2003gd,Padmanabhan:2014jta,Kolekar:2011gw,Padmanabhan:2013xyr,Jacobson:1995ab,Eling:2006aw} the dynamical equations of GR show similarity to the equations governing thermodynamic and fluid-dynamic systems. It is widely believed that in the absence of quantum theory of gravitation, the thermodynamic and the fluid-dynamic aspects of gravity may provide us with some alternative approaches/viewpoints to understand gravity in a more deeper sense.

As already mentioned above, despite huge successes of Einstein's GR, theories with modifications of  GR appears to be important to incorporate various situations, like strong gravitational regime, present observational data, etc. Among them scalar-tensor (ST) theory of gravity is one of the popular ones and attracted a huge attention \cite{Callan:1985ia,Esposito-Farese:2003jfi,Elizalde:2004mq,Saridakis:2016ahq,Crisostomi:2016czh,Langlois:2017dyl}. The theory is described by both metric tensor $g_{ab}$ and a scalar field $\phi$ which is non-minimally coupled with Ricci scalar $R$, known as Jordan frame description. The gravitational action is given by
\begin{equation}
\mathcal{S}_J = \int d^4x\sqrt{-g}L = \frac{1}{16\pi}\int d^4x\sqrt{-g}\Big(\phi R - \frac{\omega(\phi)}{\phi}\nabla_a\phi\nabla^a\phi - V(\phi)\Big)~.
\label{A1}
\end{equation}
In the above $\omega(\phi)$ is the Brans-Dicke parameter and $V(\phi)$ is an arbitrary potential. Such an action can be expressed in minimally coupled structure by using the following transformations:
\begin{equation}
g_{ab} \rightarrow \tilde{g}_{ab} = \Omega^2g_{ab};~\textrm{with}~\Omega = \sqrt{\phi}~, 
\label{A2}
\end{equation} 
and 
\begin{equation}
\phi \rightarrow \tilde{\phi} \text{ with } d\tilde{\phi} = \sqrt{\f{2\omega(\phi) + 3}{16\pi}}\f{d\phi}{\phi}~.
\label{A3}
\end{equation}
This is known as Einstein frame description and in this case the action takes the form
\begin{equation}
\mc{S}_{\text{EF}} = \int d^4x \sqrt{-\tilde{g}} \tilde{L}= \int d^4x \sqrt{-\tilde{g}}\text{ }\left(\f{\tilde{\mc{R}}}{16\pi} - \f{1}{2}\tilde{\nabla}_i\tilde{\phi}\tilde{\nabla}^i\tilde{\phi} - U(\tilde{\phi})\right)~,
\label{A4}
\end{equation}
with the non-zero potential $U = V(\phi)/ 16\pi\phi^2$. The Lagrangians (densities) in both the frames are related to each other by a total derivative term: $$\sqrt{-\tilde{g}}\tilde{L} = \sqrt{-g}L - (3/16\pi) (\sqrt{-g}\Box\phi),$$ which is unimportant as far as dynamics are concerned. Consequently, the equations of motion for metric and scalar field appears to be equivalent in both the frames (a details discussion can be followed from \cite{Faraoni:1999hp,Bhattacharya:2017pqc,Bhattacharya:2018xlq,Bhattacharya:2020jgk,Bhattacharya:2022mnb}). Despite this mathematical equivalence, it appears that certain physical properties in different frames disagree with each other, while others agree (see, e.g. \cite{Steinwachs:2011zs,Kamenshchik:2014waa,Banerjee:2016lco,Pandey:2016unk,Ruf:2017xon,Bhattacharya:2017pqc,Bhattacharya:2018xlq,Bhattacharya:2022mnb,Bhattacharya:2020wdl,Dey:2021rke}. Therefore it remains a ``dilemma'' to give preference to a particular frame \cite{Faraoni:1999hp,Faraoni:1998qx,Faraoni:2010yi}.

Recently, a series of papers \cite{PhysRevD.103.L121501, PhysRevD.104.124031, faraoni2022scalar} appear in the literature which treat the ST theory in terms of the thermodynamics of an effective scalar $\phi$-fluid. The original discussion \cite{PhysRevD.103.L121501, PhysRevD.104.124031} was done in Jordan frame where the energy-momentum (EM) tensor for fluid has been identified by comparing the equation of motion for metric tensor with that in non-vacuum GR theory. Such a tensor depends on the higher derivative of scalar field and so can be compared with that of a dissipative fluid. The authors constructed the thermodynamics of such case by using Eckart's first order thermodynamic formalism \cite{PhysRev.58.919} for a dissipative fluid. The central idea about this investigation is as follows: it is known that (vacuum) GR corresponds to ``equilibrium thermodynamic state'' as it does not incorporate any dissipative term when it is viewed as an emergent theory from a thermodynamic description \cite{Jacobson:1995ab}. While a modified theory of gravity is usually described through dissipative phenomenon (e.g. see \cite{Eling:2006aw} for $f(R)$ gravity). So usually the modified gravity theories belongs to the non-equilibrium thermodynamical state. Hence there must exists a thermodynamic limit which describes how a modified theory propagates to GR. In this formalism it is found that vanishing of temperature of $\phi$ fluid leads to GR equilibrium state \cite{PhysRevD.104.124031}. Following this idea the same set of authors investigated the thermodynamics of scalar fluid in Einstein's frame as well \cite{faraoni2022scalar}. Their approach inherently under the inequivalent picture and therefore the EM tensor in the latter case is different from that in Jordan frame. They showed that in Einstein frame GR limit is obtained by imposing vanishing of chemical potential. These investigations naturally delivers a question to ask - whether such an aspect of ST theory can have frame independent description? If so then a universal limit can be fixed to approach at GR equilibrium description. In this paper we aim to investigate this possibility through a general approach to the thermodynamics of relativistic fluids. In this regard we mention that earlier analysis was done within Eckart's formalism. Unfortunately this one suffers from acausal nature. An important improvement has been done by Israel and Stewart \cite{Warner,israel1979transient} to make the theory consistent with causality. The defined parameters in both the formalisms differ by terms, related to the viscous and heat flux of EM tensor. However for an ideal fluid EM tensor these two formalisms coincide and hence we do not need to worry about such issue. Therefore for our present case, as the EM tensor will be as that of an ideal fluid, we can use any formalism without loss of any physical restriction. 

In order to have an equivalent description the central aim will be to identity an effective scalar field fluid EM tensor which satisfies the usual relation (i.e. $\tilde{T}_{ab}^{(\tilde{\phi})} = T_{ab}^{(\phi)}/\phi$) under the transformations (\ref{A2}) and (\ref{A4}). Naturally, earlier way to read-off such quantity can not fulfil the requirement. So we adopt a different way to find the fluid description. Within various aspects of ST theory, the thermodynamic description of gravity on a Killing horizon (e.g. event horizon of a stationary black hole) appears to be equivalent in Jordan and Einstein frames (see \cite{Koga:1998un} for initial attempt and \cite{Bhattacharya:2018xlq} for improved approach). The geometrical definitions of thermodynamic entities (like temperature, energy, entropy, work term) in one frame boils down to their counter parts in other frame by the transformations (\ref{A2}) and (\ref{A3}). Spirited by this fact a thermodynamic description of field equation for metric tensor on a generic null surface (generated through normal as well as tangent vector $l^a$) has been investigated within the equivalent picture. A particular projection of this equation together with a virtual displacement along the auxiliary null vector captures a thermodynamic identity in both the frames which is of the form
\begin{equation}
\int_{S_t} d^2x T\delta_\lambda s = \delta_\lambda E + \delta_\lambda W~,
\label{A5}
\end{equation}
where $T, s, E$ are temperature, entropy density, energy respectively. $S_t$ denotes the cross section of the null surface and $\lambda$ parameterises the auxiliary null vector $k^a$ (for details, see \cite{Dey:2021rke}; for a review on this topic, see \cite{Bhattacharya:2022mnb}). The last one is the work term and is defined through $\delta_\lambda W = - \int_{S_t}\sqrt{\tilde{q}}\delta\lambda \tilde{T}^{(\tilde{\phi})}_{ab}\tilde{l}^a\tilde{k}^b$ in Einstein frame {\footnote{Here we denote tilde variable as Einstein frame quantities and without tilde ones as Jordan frame quantities. This notation, wherever necessary, will be followed.}}. Here $\tilde{T}^{(\tilde{\phi})}_{ab}$ is given by
\begin{equation}
\tilde{T}_{ab}^{(\tilde{\phi})} = \tilde{\nabla}_a\tilde{\phi}\tilde{\nabla}_b\tilde{\phi} - \f{1}{2}\tilde{g}_{ab}\tilde{\nabla}_i\tilde{\phi}\tilde{\nabla}^i\tilde{\phi} - \tilde{g}_{ab}U(\tilde{\phi})~.
\label{A6}
\end{equation}
In this equivalent description the same work term appears in Jordan frame as well and in this case the energy-momentum tensor for scalar field takes the form
\begin{equation}
\label{ogeqn}
 T_{ab}^{(\phi)} =\phi\tilde{T}_{ab}^{(\tilde{\phi})}= \phi\Big[\left(\f{2\omega + 3}{16 \pi}\right)\left\{\nabla_a\left(\ln{\phi}\right)\nabla_b\left(\ln{\phi}\right) - \f{1}{2}g_{ab}\nabla_i\left(\ln{\phi}\right)\nabla^i\left(\ln{\phi}\right)\right\} - g_{ab}\f{V(\phi)}{16\pi\phi}\Big]~.
\end{equation}
It may be noted that upto an overall multiplicative factor $\phi$, (\ref{A6}) and (\ref{ogeqn}) are same under the transformations (\ref{A2}) and (\ref{A3}). 

Now note that the work term in the thermodynamic identity is usually defined through the external pressure $P = - T^{(m)}_{ab}l^ak^b$ (see discussion in \cite{Chakraborty:2015aja,Dey:2020tkj}), where $T^{(m)}_{ab}$ is the EM tensor for the external matter field as it contributes in the work term. Therefore following the above discussion it is natural to think (\ref{A6}) or (\ref{ogeqn}) corresponds to $\phi$ fluid in ST theory. Interestingly, such an analogy can be well supported by following the fluid interpretation of dynamical equation of ST theory.  It is well known that in GR a specific projection of Einstein's equation of motion on a generic null surface yields Naiver-Stokes like equation \cite{Price:1986yy,Parikh:1997ma,Gourgoulhon:2005ng,Padmanabhan:2010rp,Kolekar:2011gw}. The same has been investigated in ST theory as well \cite{Bhattacharya:2020wdl}. It has been observed that if one demands an equivalent picture for fluid variables in both the frames then (\ref{A6}) and (\ref{ogeqn}) provides the external forcing terms in Einstein and Jordan frames respectively (see \cite{Bhattacharya:2020wdl,Bhattacharya:2022mnb} for details). Looking at this facts we consider (\ref{A6}) or (\ref{ogeqn}) as the EM tensor for the $\phi$ fluid in ST theory. Moreover, these forms are equivalent and applied in both the frames. Therefore a thermodynamic description of this fluid captures the sprit of the equivalent picture. 

It is worth to mention that in \cite{Dey:2021rke}, the authors discussed the thermodynamic interpretation in Einstein frame in which their approach identifies the EM tensor as (\ref{A6}). Therefore according to our above discussion such analysis can be considered as an equivalent description. However we have few observations on the previous analysis which need to be further investigated. We mention them in the next discussion. This will formulate the basis of our present investigation and hopefully will lead to a fruitful description of thermodynamics of $\phi$ fluid in a frame independent setup. 

Therefore, in order to setup the plot, in next couple of sections, we first briefly discuss the thermodynamics of relativistic fluids and then mention about our observations on the previous works \cite{PhysRevD.103.L121501, PhysRevD.104.124031, faraoni2022scalar}. In the third section, we present the present status of ST theories via fluid thermodynamic viewpoints. In section 4, in order to further proceed towards our goal, we motivate the reader towards a new EM tensor. In the next section, we develop a general formalism to describe a mapping between a minimally coupled scalar field and corresponding fluid/thermodynamic quantities. Using the same mapping the fluid/thermodynamic quantities for our case in both the frames are being found in this section as well. Finally, in section 6, we conclude the paper with some future directions. 

\section{Summary of the thermodynamics of relativistic fluids}\label{Sec2}
Consider a fluid with four-velocity $u^a$ satisfying $u_au^a = -1$ moving on a spacetime manifold $(\mathcal{M}, g_{ab}, \nabla)$. We choose the fluid velocity along time direction and foliate the spacetime into (spacelike) hypersurfaces orthogonal to flow lines. The projector onto these hypersurfaces as well as the induced metric is then given by
\begin{equation}
h_{ab} = g_{ab} + u_au_b~.
\label{B1}
\end{equation}
The acceleration $\dot{u}_a = u^b\nabla_bu^a$ is perpendicular to the velocity i.e. $\dot{u}_au^a = 0$. In general a viscous fluid is described the following energy-momentum tensor
\begin{equation}
\label{emfull}
T_{ab} = \underbrace{\rho u_au_b + p_{\text{iso}}h_{ab}}_{T_{ab}^{\text{ideal}}} + \underbrace{q_au_b + q_au_b}_{T_{ab}^{\text{heat}}} + \underbrace{p_{\text{vis}}h_{ab} + \pi_{ab}}_{T_{ab}^{\text{vis}}}~,
\end{equation}
which is covariantly conserved; i.e. $\nabla_a T^{ab} = 0$.
Here $\rho, p_{\text{iso}}, q_a, p_{\text{vis}}$ and $\pi_{ab}$ are energy density, isotropic pressure, heat flux, viscous pressure and traceless part of the viscous stress-tensor (i.e. $\pi^a_a=0$), respectively. $q_a$ is chosen to be spatial vector; i.e. $u^aq_a=0$ and also we have $u^a\pi_{ab}=0$.
Choosing the viscous stress tensor is linear in the velocity gradients and proportional to the (scalar) coefficient of viscosity $\eta$ the suggested form for $\pi_{ab}$ is given by 
\begin{equation}
\label{piab}
\pi_{ab} = - \eta \left\{h_{a p}h_{b q} \left(\nabla^{q}u^{p} + \nabla^{p}u^{q}\right) - \f{2}{3}h_{ab}\nabla_{p}u^{p}\right\} = - 2 \eta \sigma_{ab}~. 
\end{equation}
In the above $\sigma_{ab}$ is known as shear.
The ``bulk viscosity'' $\zeta$ is defined through $p_{\text{vis}}$ which is related to the trace part of the viscous stress-tensor:
\begin{equation}
\label{pvis}
p_{\text{vis}} = - \zeta \theta~, 
\end{equation}
where $\theta = \nabla_au^a$, called as expansion.
Apart from the conservation energy-momentum tensor (\ref{emfull}), the fluid dynamics is supplemented by conservation of mass current ($\nabla_am^a = 0$) and conservation of total number current ($\nabla_a N^a=0$), where $m^a = mu^a$ with $m = \sqrt{-m_am^a}$ is the mass density and
\begin{equation}
\label{Na}
N^a = n^a + \nu^a~.
\end{equation}
In the above one has $n^a = nu^a$ and $\nu^a$ is known as diffusion flux. By construction $u_a\nu^a = 0$ i.e. the diffusion flux $\nu^a$is spatial, so that the number density can be obtained as $n = - u_aN^a$. The total number current is conserved i.e. $\nabla_aN^a = 0$, so we can define the fluid chemical potential $\mu$. 

An approach to understand the (first order) thermodynamics of an imperfect fluid is using Eckart's formalism \cite{PhysRev.58.919}. This formalism provides a relation between the thermodynamic parameters and the fluid parameters under the input that the viscous fluid satisfies usual laws of thermodynamics. Although originally the formalism was developed in a flat background we work in some arbitrary curved spacetime with metric signature $(- + + +)$. Also we will include particle exchange which was missing in the original analysis. The whole generalization has been done in \cite{andersson2007relativistic}. Here, we will briefly state the results. The entropy current in the system can be written as $s^a = su^a + \beta q^a - \lambda \nu^a$
where $s$ is the entropy density and $\beta,\lambda$ are unknown parameters. These parameters are fixed by imposing the second law $\nabla_as^a \geq 0$ along with the use of thermodynamic identity (written in per unit volume)
\begin{equation}
\nabla_a\rho = T\nabla_as + \mu \nabla_an~, 
\end{equation}
and Euler's relation
\begin{equation}
\rho + p_{\text{iso}} = Ts + \mu n~.
\label{Euler}
\end{equation}
Here $T$ is the temperature of the fluid.
Such imposition fixes those unknown parameters as
\begin{equation}
\beta = \f{1}{T}~; \,\,\,\,\
\lambda = \f{p+\rho}{nT} - \f{s}{n} = \f{\mu}{T}~,
\end{equation}
along with a particular choices of $q_a$ and $\nu_a$.
Thus, the salient features of {\it extended} - Eckart's formalism are summarised as follows: 
\begin{align}
q^a &= - \kappa h^{ab}\left(\nabla_bT + \dot{u}_bT\right)~; 
\label{qeck}
\\ 
\pi^{ab} &= - 2 \eta \sigma^{ab}~;
\label{pieck} 
\\ 
p_{\text{vis}} &= - \zeta \theta~;
\label{pviseck} 
\\
\nu^a &= - \sigma T^2h^{ab}\nabla_b\left(\f{\mu}{T}\right)~. 
\label{nueck}
\end{align}
Here $\kappa$ is heat conductivity and $\sigma$ is known as diffusion coefficient. Also remember that $\kappa, \sigma \geq 0$.
The above ones relate the thermodynamic parameters of a viscous fluid with its fluid parameters (the details are given in \cite{andersson2007relativistic}).
In the original construction \cite{PhysRev.58.919}, the bulk viscosity $\zeta$ was taken to be zero and the velocity $u^a$ was taken to be the velocity of the matter i.e. the fluid. This is a particular choice of a local rest frame (the Eckart frame), and in this choice $\nu^a = 0$. If we choose a particular rest frame, we may have to impose conditions on the above quantities which arise from the physical restrictions of the rest frames. However, as of yet the above description and (\ref{qeck})-(\ref{nueck}) are much more general.

Although Eckart's formalism is a fundamental approach to studying the thermodynamics of relativistic fluids, the theory has limitations since it does not account for causality and exhibits instability issues. Extensions of this theory, such as those by Israel and Stewart \cite{Warner,israel1979transient}, aim to fix these issues. The Israel - Stewart formalism is then a stable, causal model to study relativistic fluids. In this approach, the fundamental fluid quantities (\ref{qeck})-(\ref{nueck}) are modified to include second order perturbations. This modification comes from terms describing dissipation in the fluid itself. In principle, one must use this more general approach when studying relativistic fluids. However in the absence of dissipation i.e. when studying ideal fluids, Eckart's formalism is sufficient as both the formalisms coincide in this limit. In our analysis we consider an EM tensor which has ideal fluid structure (see Eqs. (\ref{A6}) and (\ref{ogeqn})). Therefore although the explicit relationship between temperature and the dissipative quantities may be modified by using the Israel - Stewart formalism (see \cite{israel1979transient}), in our analysis these modifications will vanish since they will depend on dissipative quantities which are absent in our problem. Therefore, the two approaches yield identical results and hence, relations (\ref{qeck}) - (\ref{nueck}) can be used in this context.

\section{Status quo of scalar-tensor gravity via fluid thermodynamics}
In a series of recent papers by Faraoni et al \cite{PhysRevD.103.L121501,PhysRevD.104.124031,faraoni2022scalar}, scalar-tensor gravity is described through the thermodynamics of a fluid, employing Eckart's first order thermodynamics formalism \cite{PhysRev.58.919}. In these papers, they use Eckart's formalism to consider scalar-tensor gravity both in the Jordan and the Einstein frame, and provide one approach to reconciling the differences between the two frames. However, this analysis requires a few inputs and logistics which may be put under scanner. We elucidate them below. 
\begin{enumerate}
\item Fundamental ``discrepancy'' between the fluid energy-momentum tensors:  
Beginning with the appropriate action in either frame, the authors write down the field equations in such a way so as to treat any new terms apart from the Einstein tensor in the field equations to be a part of the fluid energy-momentum tensor. That is, the field equations are written as, 
$G_{ab} = T_{ab}^{(\phi)}$ and $\tilde{G}_{ab} = \tilde{T}_{ab}^{(\tilde{\phi})}$, where we use tilde variables to indicate the Einstein frame quantities. Here, $G_{ab}$ is the usual Einstein tensor and any extra terms which appear due to the terms apart from the Einstein Hilbert action, are grouped into $T^{(\phi)}_{ab}$. However, this naively motivated way of identifying an EM tensor for the fluid lacks any physically motivated argument and more over the EM tensor in both the frames appears to be inequivalent i.e. they are not related by the transformations (\ref{A2}) and (\ref{A3}) between the Einstein and the Jordan frames.
Furthermore, this inequivalence manifests into the thermodynamic description that follows, as we will point out in the next discussion. 
\item Absence of a single thermodynamic description between the frames: 
In \cite{PhysRevD.103.L121501, PhysRevD.104.124031}, the authors discuss scalar-tensor gravity in the Jordan Frame. The aforesaid way of identification of EM tensor for $\phi$ fluid incorporates viscous part. In this process, they obtain a non-zero heat flux \cite{pimentel1989energy, PhysRevD.98.084019}, from which they eventually obtain the ``temperature of gravity" using Eckart's heat flux. However, a crucial assumption used in this analysis is the vanishing spatial temperature gradient i.e. $h^{ab}\nabla_bT = 0$, which is used to find out the specific conductivity and the temperature, and ultimately show that the GR limit corresponds to a $T = 0$ state for the fluid. In this analysis, the number density $n$, the chemical potential $\mu$ and the entropy density $s$ are not discussed and therefore they are not computed. 

    However, in \cite{faraoni2022scalar}, the authors apply Eckart's formalism and use the existing dictionary between perfect fluids and minimally coupled scalar fields in \cite{piattella2014note} to study the Einstein frame description of scalar-tensor gravity. Here, they demand that the temperature must vanish identically, for Eckart's formalism to be consistent with the vanishing heat flux for the perfect fluid. Now to provide a mechanism in obtaining GR as some stable thermodynamic limit, they introduce the chemical potential $\mu$ which leads to a vanishing diffusion flux 
\begin{equation}
\label{faraoninu}
\nu^a = - \mc{D}h^{ab}\left[\nabla_b\mu + \dot{u}_b \mu\right] = 0~,
\end{equation}
where $\mc{D}$ is the diffusion coefficient.
Inspired by Eckart's heat flux (\ref{qeck}) and the work in \cite{kremer2014diffusion}, the authors chose the above form for $\nu^a$. Note that this form is different from (\ref{nueck}) by the last term. The chemical potential is such that the GR limit corresponds to vanishing chemical potential. Thus, GR limit corresponds to $T = 0$ in Jordan frame, while $\mu = 0$ in Einstein frame.
    
    The above two points, therefore, indicates an inequivalent description of thermodynamics between the two frames.
    
\item Ambiguity in the choice of $\nu^a$: Within this inequivalent description a comment can be made on the choice (\ref{faraoninu}) for the diffusion flux.
This form of the diffusion flux does not guarantee consistency with the second law of thermodynamics except Eckart's frame in which $\nu^a=0$. In calculating the divergence of the entropy current $\nabla_as^a$, the contribution due to the diffusion flux is given by $- \nu^a\nabla_a\lambda$ (see Eq. (282) of \cite{andersson2007relativistic}). Then for the choice (\ref{faraoninu}) it yields
\begin{equation}
- \nu^a\nabla_a\lambda= \f{\nu^a\nu_a}{\mc{D}} + \mc{D}\mu\nu^a\dot{u}_a~.
\end{equation} 
While the first term in the right hand side is always positive since $\nu^a$ is spacelike, the second term is ambiguous. Although the two vectors are individually spacelike, the contraction may not be positive. While a vanishing $\nu^a$ can be considered (as done in Eckart's frame \cite{andersson2007relativistic}), this also may face an issue as described below, particularly for perfect fluid. Moreover, this choice dose not guarantee the positivity of $\nabla_a s^a$ in other frames.

\item Eckart frame and a vanishing diffusion current:
Conventionally, choosing the Eckart frame implies the velocity is taken to be along the flow of the fluid particles. Thus, the diffusion current must vanish i.e. $\nu^a = 0$. With the choice of $\mu$ in \cite{faraoni2022scalar} (i.e. Eq. (\ref{faraoninu})) this is indeed the case. However, the vanishing of the diffusion flux places restrictions on the theory under consideration, particularly in Einstein's frame. In the Einstein frame, if one considers the number current defined naively as $n u^a$, then it is no longer covariantly conserved due to the dynamics of the field $\phi$. Consider an action of the form 
\begin{equation}
\mc{S} = \int \sqrt{-g}\text{ }d^4x\text{ }\mc{L}\left(\psi,X\right)~, 
\label{5.1}
\end{equation}
with $X=-(1/2) \nabla_i\psi\nabla^i\psi$~. The Einstein frame ST gravity action is an example of this form of action. For this class of actions, the energy momentum tensor is given by,
\begin{equation}
T_{ab} = - \f{2}{\sqrt{-g}}\f{\delta \mc{S}}{\delta g^{ab}} = \mc{L}_X \nabla_a\psi \nabla_b\psi + \mc{L} g_{ab} ~,
\label{tabapp}
\end{equation}
and is covariantly conserved i.e. $\nabla^aT_{ab} = 0$. Then, we can write down the Euler Lagrange equation for the field $\psi$ as,
\begin{equation}
\nabla_a\left(nu^a\right) = - \mc{L}_{\phi}~,
\label{3.3}
\end{equation}
where $\mc{L}_{\phi} = \p \mc{L}/\p \phi$. The chemical potential defined in this context is ambiguous. This can be remedied by modifying the number current to include the diffusion flux \cite{andersson2007relativistic}, such that $N^a = nu^a + \nu^a$ with $\nabla_aN^a = 0$. Thus one finds
\begin{equation}
\label{ndifcond}
\nabla_a\left(nu^a\right) = - \nabla_a\nu^a = - \mc{L}_{\phi}~,
\end{equation}
and therefore imposition of $\nu^a = 0$ leads to $\mc{L}_{\phi} = 0$. This suggests that the Lagrangian must be independent of $\phi$. In other words, the conservation of $nu^a$ is connected to the fact that the Lagrangian must be independent of $\phi$. This type of Lagrangian usually must be independent of potential term and then we could have a Lagrangian that only contains the kinetic terms involving $X$. However, ST theory in the Einstein frame in a general framework is given by the action (\ref{A4}). This gives an energy momentum tensor for $\phi$ as (\ref{A6}).
Note that this one resemblances to that of a perfect fluid.
In that case if $\mc{L}_{\phi} = 0$, then the Lagrangian being independent of $\phi$ would imply that the above potential term in the Lagrangian must not exist. It would refer to describing a minimally coupled free scalar field, in curved space-time. Hence, this is incompatible with considering a general ST theory in the Einstein frame. 
    
Observe that we obtain this restriction on Lagrangian directly comes from imposition of $\nu^a = 0$ which amounts to choosing the Eckart frame. Thus it seems that, {\it in general to describe minimally coupled scalar fields whose EM tensor is identical to that of a perfect fluid, the Eckart frame is not the perfect choice to describe its thermodynamics}. 

  \end{enumerate}

Thus, the description presented so far has a few limitations and needs to revisited so as to obtain an equivalent picture, from a thermodynamic and gravitational perspective. 

\section{Motivating a new EM tensor}

We want a description of the thermodynamics of ST gravity that is equivalent between the two frames, so as to provide such an interpretation for the gravity side. To do so, we first realise that the scalar field in ST theory contributes to the description of gravity, as well as to describe the $\phi$ fluid. We adopt the approach used in \cite{Dey:2021rke,Bhattacharya:2022mnb} where we use the fact that projecting the gravitational field equations onto a null surface can give us an equation similar in structure to the first law, from which we identify an EM tensor such that the individual thermodynamic quantities between the two frames are in correspondence. Doing so provides us with the EM tensor (\ref{ogeqn}). 
The details of the derivation of this are given in \cite{Dey:2021rke}. Such an identification, as mentioned in the introduction, is also bolstered from fluid description of ST gravitational theory in an equivalent platform \cite{Bhattacharya:2020wdl}. Interestingly, such EM tensor upto an over all factor boils down to (\ref{A6}) under the transformations (\ref{A2}) and (\ref{A3}) and moreover the Einstein frame EM tensor structure can be derived from the last two terms of (\ref{A4}) and hence is covariantly conserved ($\tilde{\nabla}^a\tilde{T}_{ab}^{(\tilde{\phi})} = 0$). What is important is that this is the EM tensor that emerges in \textit{both} frames if we define it keeping in mind thermodynamic equivalence of the gravitational theory; any analysis performed using this $\tilde{T}_{ab}^{(\tilde{\phi})}$ holds in both frames. Thus, relying on thermodynamic equivalence naturally leads with a hint of gravitational equivalence, at least at the level of the EM tensor. Further, note that the structure of (\ref{A6}) is similar to that of a perfect fluid. Since this way of identified EM tensor is same in both the frame, it suffices to perform all our analysis in the Einstein frame description. Also note that the action for this EM tensor can be considered as the last two terms in (\ref{A4}) which falls within the general class of actions considered in \cite{faraoni2022scalar}. However, performing the same mapping will still not be useful, even with this modified $T_{ab}^{(\phi)}$ due to a vanishing diffusion flux. As mentioned earlier, since $\nabla_a\nu^a = \mc{L}_{\phi}$, a vanishing $\nu^a$ would mean that the Lagrangian is independent of $\phi$. Thus, the Eckart frame may not be the ideal choice to describe a perfect fluid. 

In the following section we will develop a general formalism to describe a mapping between a minimally coupled scalar field and corresponding fluid/thermodynamic quantities. This can then be used to study the particular example of the $\phi$ fluid in the Einstein frame ST theory. We will then discuss how on can also obtain the Jordan frame picture without additional work. 

\section{A possible alternate way out}
We continue to consider the general class of actions for a minimally coupled scalar field, since it is sufficient to demonstrate this approach at least in the Einstein frame. Moreover, our identified tensor (\ref{A6}) belongs to this class as well.
Therefore consider a minimally coupled scalar field $\psi$, with a general Lagrangian density $\mc{L}$ whose action is (\ref{5.1}). 
%\begin{equation}
%\mc{S} = \int \sqrt{-g}\text{ }d^4x\text{ }\mc{L}\left(\psi,X\right) \hspace{7 px}\text{;}\hspace{7 px} X = -\f{1}{2}\nabla_i\psi\nabla^i\psi
%\label{5.1}
%\end{equation}
The energy momentum tensor for this field can be derived as (\ref{tabapp})
%\begin{equation}
%\label{gentab}
%T_{ab} = -\f{2}{\sqrt{-g}}\f{\delta \mc{S}}{\delta g^{ab}} = \mc{L}_X\nabla_a\psi\nabla_b\psi + g_{ab}\mc{L}~,
%\end{equation}
with the notation $\mc{L}_X = \p \mc{L}/\p X$ and $\mc{L}_{\psi} = \p \mc{L}/\p \psi$. Note that this $T_{ab}$ is conserved. 

Comparing (\ref{tabapp}) with (\ref{emfull}), we have the following preliminary dictionary: 
\begin{equation}
\label{dict1}
p =p_{\text{iso}}= \mc{L}~;\,\,\,\ u_a = \f{\nabla_a\psi}{\sqrt{2X}}; \,\,\,\ \rho = 2X\mc{L}_X - \mc{L}~.
\end{equation}
We also see that there is no term corresponding to a heat flux or viscous stress i.e. $p_{\text{vis}} = 0$, $\pi_{ab} = 0$ and $ q_a = 0$. This is reasonable because we do not expect a minimally coupled scalar field to generate any dissipation. Thus, the scalar field is mapped to a perfect fluid. Now from thermodynamics, the enthalpy is defined as $h = \f{\rho + p}{n}$.
We choose the flow potential normalisation through the enthalpy as done in \cite{piattella2014note}. That is to say, we choose,  
\begin{equation}
hu_a = \nabla_a\psi~.
\end{equation}
Then (\ref{dict1}) yields
\begin{equation}
\label{dict2}
n = \sqrt{2X}\mc{L}_X~.
\end{equation}

So far, what we have mentioned about the fluid variables in terms of field $\psi$, is exactly identical to that in \cite{faraoni2022scalar}. If we now look at the dynamics of $\psi$ we might be able to describe the dynamics of the fluid through our growing dictionary. From the dynamical equations of the field, we have, $\nabla_a\left(\mc{L}_X\nabla^a\psi\right) = \nabla_a\left(nu^a\right) = - \mc{L}_{\psi}$. Thus, we see that particle number is not conserved in this fluid. This suggests that there is some diffusion, and our definition of the chemical potential for the fluid must be connected to this diffusion. Such a situation has been taken into account in \cite{andersson2007relativistic} to provide a general version of Eckart's formulation. This we already mentioned in Section \ref{Sec2}. In this case the number current is given by (\ref{Na}). 
Then conservation it yields
\begin{equation}
\label{divnu}
\nabla_a\nu^a = - \nabla_a\left(nu^a\right) = \mc{L}_{\psi}~.
\end{equation}
Moreover, we can now follow the dictionary (\ref{qeck}) -- (\ref{nueck}) to find the thermodynamic quantities of our scalar fluid.
Then since the heat flux $q^a$ vanishes, we have 
\begin{equation}
\label{eckartq}
q^a = -\kappa h^{ab}\left[\nabla_b T + \dot{u}_b T\right] = 0~.
\end{equation}
To satisfy the above criteria, in \cite{faraoni2022scalar} $T=0$ has been chosen. If we choose $T = 0$, then the present choice of $\nu^a$ (given by Eq. (\ref{nueck})) yields
\begin{equation}
\nu^a = \sigma \mu h^{ab}\nabla_bT - \sigma Th^{ab} \nabla_b\mu = 0~.
\label{5.8}
\end{equation}
Remember that even the choice (\ref{faraoninu}), adopted in \cite{faraoni2022scalar}, also implies vanishing of $\nu^a$.
But as discussed earlier $\nu^a = 0$ with (\ref{divnu}) implies that the Lagrangian is independent of $\psi$, which is in contradiction to the present situation as the Lagrangian explicitly depends of $\psi$. Moreover in Eckart's frame one has $\nu^a=0$. Therefore this discussion seems to indicate that the present system can not be well described in this frame and $T=0$ may not be a good choice. Hence we propose an alternative choice for $T$ to satisfy (\ref{eckartq}).
In order to work in the paradigm of equilibrium thermodynamics, we demand that the temperature be constant and positive definite. Hence, we take $T = T_0$, a constant and $T_0>0$. From (\ref{eckartq}), this would mean that, 
\begin{equation}
- \kappa \dot{u}^a T_0 = 0~.
\end{equation}
We choose $\kappa=0$; i.e. vanishing thermal conductivity. This would mean that we consider a fluid which is at constant temperature, but is a perfect insulator. This can be considered as a consequence of the fact that we have a perfect fluid which does not allow a heat flux. 
However, with a non-zero temperature, we can proceed to define a non-vanishing diffusion flux. Since the temperature is a constant, (\ref{nueck}) implies
\begin{equation}
\label{newnu}
\nu^a =  - \mc{D}h^{ab}\nabla_a\mu~,
\end{equation}
where $\mc{D} = \sigma T$ is taken to be the (scaled) diffusion coefficient. Note that this is the first term in the choice (\ref{faraoninu}).

With this choice, we now want to find a closed form expression for the chemical potential. To do so, we try to set up a differential equation for the chemical potential through the Euler relation (\ref{Euler}). A simple computation yields
\begin{equation}
h^{ab}\nabla_a\mu =  h^{ab}\nabla_a\left(\sqrt{2X}\right)~.
\label{5.14}
\end{equation}
A detailed derivation of the above equation is presented in Appendix \ref{AppB}.
A solution for $\mu$ can be taken as
\begin{equation}
\label{dict3}
\mu = \sqrt{2X} - f(\psi)~.
\end{equation}
for some arbitrary $f$ which is function of $\psi$ only. Using the facts $\nabla_a \psi \propto u_a$ and $h^{ab}u_a = 0$ one can check that it satisfies (\ref{5.14}). For $f = 0$ this yields $\mu = \sqrt{2X}$ which was obtained in \cite{faraoni2022scalar}. But in the present case we can not choose $f=0$ as this is not consistent with a non-zero temperature as shown below. When $f=0$ then we have $\mu = \sqrt{2X}$ and then the Euler relation (\ref{Euler}) along with (\ref{dict1}) and (\ref{dict2}) imply $Ts=0$.
Therefore, when $\mu = \sqrt{2X}$ and $s \neq 0$, we must have $T = 0$. Hence, we instead determine the chemical potential upto the function $f(\psi)$, given by (\ref{dict3}).
Also note that the chemical potential, depending on the values of $\sqrt{2X}$ and  $f$, can be positive or negative.
Then the Euler relation (\ref{Euler}) fixes the entropy density as
\begin{equation}
s = \f{\sqrt{2X}\mc{L}_X}{T} f(\psi) = \f{\sqrt{2X}\mc{L}_X}{T_0} f(\psi)~. 
\end{equation}
To ensure that the entropy density is positive, we impose $f>0$. 

Finally, from the divergence of the diffusion, one can find the diffusion constant $\mc{D} = \sigma T_0 $. This is done using equation (\ref{newnu}). This yields
\begin{equation}
\nabla_a\nu^a = -\nabla_a\left(\mc{D} h^{ab}\nabla_b\mu\right) = -\nabla_a\left[\mc{D} h^{ab}\nabla_a \left(\sqrt{2X}\right)\right] = \mc{L}_{\psi}
\end{equation}
where the last equation is obtained using (\ref{divnu}). Then the diffusion coefficient $\mc{D}$ at least formally be obtained as $\sigma(\psi,X) = \mc{D}(\psi,X)/T_0$ by solving the above equation. 
These results are summarised in Table 
\ref{dictionary}. 

\begin{table}[ht!]
    \centering
    \begin{tabular}{c c } \toprule 
    Fluid and thermodynamic Variable   & Field Theory Variable \\ \midrule 
    $X$ & $\left(-1/2\right)\nabla_a\psi\nabla^a\psi$ \\ 
    $p$ & $\mc{L}$ \\
    $u_a$ & $\nabla_a \psi/\sqrt{2X}$ \\ 
    $\rho$ & $2X\mc{L}_X - \mc{L}$ \\ 
    $n$ & $\sqrt{2X}\mc{L}_X$ \\ 
    $h$ & $\sqrt{2X}$ \\  
    $\mu$ & $\sqrt{2X} - f(\psi)$ \\ 
    $s$ & $\sqrt{2X} \mc{L}_X f(\psi)/ T$ \\ 
    $T$ & $T_0$ (\textrm{a constant}) \\ 
    \bottomrule 
    \end{tabular}
    \caption{Mapping between fluid and scalar field variables}
    \label{dictionary}
\end{table}

This dictionary is a general prescription. To work in the Einstein frame with the Einstein variables, one would simply have to let $\psi = \tilde{\phi}$ and $\nabla_a = \tilde{\nabla}_a$. All thermodynamic quantities are then computed as above with 
\begin{equation}
\tilde{\mathcal{L}} = -(1/2)\tilde{\nabla}_i\tilde{\phi}\tilde{\nabla}^i\tilde{\phi} - U(\tilde{\phi})~.
\label{5.15} 
\end{equation}
Using this Lagrangian and the Table \ref{dictionary}, one obtains for the Einstein frame quantities as shown in Table \ref{Einsteintable}.
\begin{table}[h!]
    \centering
    \begin{tabular}{c c c c c } \toprule 
    %\multicolumn{5}{c}{} \\  \midrule
    $\tilde{u}_a$ & = & ${\tilde{\nabla}_a\tilde{\phi}}/{\sqrt{2\tilde{X}}}$ & = &$ \sqrt{\phi}{\nabla_a\left(\ln{\phi}\right)}/{\sqrt{2X_J}}$ \\ 
    $\tilde{p} $& = & $ \tilde{\mc{L}} $ & = & $ \left({2\omega + 3}\right)/\left({16\pi\phi}\right)X_J - {V(\phi)}/{16\pi\phi^2}$ \\ 
    $\tilde{\rho}$ & = & $ 2\tilde{X}\tilde{\mc{L}}_{\tilde{X}} - \tilde{\mc{L}}$ & = & $ \left({2\omega + 3}\right)/\left({16\pi\phi}\right)X_J + {V(\phi)}/{16\pi\phi^2}$ \\ 
    $\tilde{n} $& = &$ \sqrt{2\tilde{X}}\tilde{\mc{L}}_{\tilde{X}} $ & = & $ \sqrt{\left({2\omega + 3}\right)/\left({16\pi \phi}\right)2X_{J}}$ \\ 
    $\tilde{h} $ & = & $ \sqrt{2\tilde{X}}$ &= &  $ \sqrt{\left({2\omega + 3}\right)/\left({16\pi \phi}\right)2X_{J}}$ \\ 
    $\tilde{\mu}$ &= &  $\sqrt{2\tilde{X}} - \tilde{f}(\tilde{\phi})$ &= &  $\sqrt{\left({2\omega + 3}\right)/\left({16\pi \phi}\right)2X_{J}} - \tilde{f}\left(\phi\right)$ \\ 
    $\tilde{s} $ &= &  $\sqrt{2\tilde{X}}\tilde{\mc{L}}_{\tilde{X}}\tilde{f}(\tilde{\phi})/\tilde{T}$ &= &  $\sqrt{\left({2\omega + 3}\right)/\left({16\pi \phi}\right)2X_{J}}~{\tilde{f}(\phi)}/{T_0}$ \\ 
    $\tilde{T}$ &= &  $\tilde{T}_0$ (\text{constant})&  &   \\ 
    \bottomrule 
    \end{tabular}
    \caption{Fluid quantities in the Einstein frame}
    \label{Einsteintable}
\end{table} 

In this table we have defined $X_J = (-1/2)\nabla_a\left(\ln{\phi}\right)\nabla^a\left(\ln{\phi}\right)$ and for the Lagrangian $\tilde{\mc{L}}$ one has $\mc{\tilde{L}}_X = 1$. The second equality in this table is obtained using the conformal transformations (\ref{A2}) and (\ref{A3}). 

We can use the dictionary given in Table (\ref{dictionary}) to determine the fluid variables in the Jordan frame as well. To do so, we first require a Lagrangian for the $\phi$ fluid in the Jordan frame. This can be achieved by the requirement that the corresponding action is same in both the frames; i.e. 
\begin{equation}
\int d^4x\sqrt{\tilde{g}}\tilde{\mc{L}} = \int d^4x \sqrt{-g}\mc{L}_J~.
\end{equation}
This yields $\mc{L}_J = \phi^2 \tilde{\mc{L}}$.
Hence from (\ref{5.15}) we have 
\begin{equation}
\mc{L}_J =  -\f{\phi}{2}\left(\f{2\omega+3}{16\pi}\right)\nabla_a\left(\ln{\phi}\right)\nabla^a\left(\ln{\phi}\right) - \f{V(\phi)}{16\pi}~.
\end{equation}
With this Lagrangian, we can also derive the EM tensor $T_{ab}^{(\phi)}$ in (\ref{ogeqn}) by considering variations with respect to $g^{ab}$, the Jordan frame metric. Therefore, we identify this as the $\phi$ fluid Lagrangian in the Jordan frame. Using this Lagrangian and the general dictionary (shown in Table \ref{dictionary} ), we can now compute the various thermodynamic quantities in the Jordan frame. This is done in Table (\ref{jordantable}) below.

\begin{table}[ht!]
    \centering
    \begin{tabular}{c c c} \toprule 
    %\multicolumn{3}{c}{} \\ \midrule 
    %$X_J$ & = & $\left(-1/2\right)\nabla_a\left(\ln{\phi}\right)\nabla^a\left(\ln{\phi}\right)$ \\ 
    $u^J_{a}$ & = & $\nabla_a \left(\ln{\phi}\right)/\sqrt{2X_J}$ \\ 
    $p_J$ & = & $\mc{L}_J$ \\
    $\rho_J$ & = & $\phi\left(2\omega + 3)/(16\pi\right)X_J + V(\phi)/16\pi$ \\ 
    $n_J$ & = & $\sqrt{2X_J}~\phi\left(2\omega + 3)/(16\pi\right)$ \\ 
    $\mu_J$ & = & $\sqrt{2X_J} - f(\phi)$\\ 
    $h_J$ & = & $\sqrt{2X_J}$ \\  
    $s_J$ & = & $\sqrt{2X_J}~\phi\left(2\omega + 3)/(16\pi\right) f(\phi)/ T_0$ \\ 
    $T_J$ & = & $T_{0} = \tilde{T}_0$ \\ 
    \bottomrule 
    \end{tabular}
    \caption{Fluid quantities in the Jordan frame}
    \label{jordantable}
\end{table}

Firstly, in our formulation, we have chosen $T_J = T_0 = \tilde{T}_0$, which is a constant. This is because in our formulation we choose temperature of scalar fluid as a constant value. Since $\tilde{T}_0$ is the temperature in Einstein frame, the same in Jordan frame should not be scaled by $\phi$; otherwise it would not be constant in the latter frame.  
Now, consider the following quantity:
\begin{align}
\tilde{X} = -\f{1}{2}~\tilde{g}^{ab}~\tilde{\nabla}_a\tilde{\phi}\tilde{\nabla}_b\tilde{\phi} = -\f{1}{2}\left(\f{2\omega + 3}{16\pi}\right)\f{g^{ab}}{\phi}\nabla_a\left(\ln{\phi}\right)\nabla_b\left(\ln{\phi}\right) = \left(\f{2\omega + 3}{16\pi\phi}\right)X_J~. 
\label{Xeqn}
\end{align}
In a similar way, we can relate the other quantities:
\begin{align}
    u^J_a &= \f{\nabla_a\left(\ln{\phi}\right)}{\sqrt{2X_J}} = \f{\tilde{u}_a}{\sqrt{\phi}}~, \label{ueqn}\\ 
    p_J &= \mc{L}_J = \phi^2\tilde{\mc{L}} = \phi^2\tilde{p}~, \label{peqn}
\end{align}
\begin{align}
    \rho_J &= \phi\left(\f{2\omega + 3}{16\pi}\right)X_J + \f{V(\phi)}{16\pi} = \phi^2~\tilde{\rho}~, \label{rhoeqn}\\ 
    n_J = \sqrt{2X_J}~\phi\left(\f{2\omega + 3}{16\pi}\right) &= \phi~\sqrt{\phi\left(\f{2\omega + 3}{16\pi}\right)}\sqrt{\left(\f{2\omega + 3}{16\pi\phi}\right)2X_J} = \phi~\sqrt{\phi\left(\f{2\omega + 3}{16\pi}\right)}~\tilde{n}~. \label{neqn}
\end{align}
Next in order to the validity of the Euler relations $\tilde{\rho} + \tilde{p} = \tilde{T}_0\tilde{s} + \tilde{\mu} \tilde{n}$ and $\rho_J + p_J = T_0 s_J +\mu_J n_J $ in the respective frames one must has $s_J = \phi^2\tilde{s}$ and $\mu_Jn_J = \phi^2\tilde{\mu}\tilde{n}$.
From Table (\ref{Einsteintable}), Table (\ref{jordantable}) and (\ref{neqn}) we have 
\begin{align}
    \mu_J~n_J &= \left\{\sqrt{2X_J} - f(\phi)\right\}\left\{\phi~\sqrt{\phi\left(\f{2\omega + 3}{16\pi}\right)}~\tilde{n}\right\} = \phi^2\left\{\sqrt{\left(\f{2\omega + 3}{16\pi\phi}\right)2X_J} - \sqrt{\left(\f{2\omega + 3}{16\pi\phi}\right)}f(\phi)\right\}\tilde{n} 
    \nonumber
 \\
    &= \phi^2\left\{\sqrt{2\tilde{X}}- \sqrt{\left(\f{2\omega + 3}{16\pi\phi}\right)}f(\phi)\right\}\tilde{n}~,
\end{align}
Now the above will satisfy $\mu_Jn_J = \phi^2\tilde{\mu}\tilde{n}$ if the functions $\tilde{f}(\phi)$ and $f(\phi)$ are related as
\begin{equation}
\label{feqn}
\tilde{f}(\phi) = \sqrt{\frac{\left({2\omega + 3}\right)}{\left({16\pi\phi}\right)}} f(\phi)~.
\end{equation}
Thus, this fixes the relationship between the functions $\tilde{f}(\phi)$ and $f(\phi)$. Then the chemical potentials are related as 
\begin{align}
    \mu_J = \sqrt{2X_J} - f(\phi) = \left(\f{2\omega + 3}{16\pi\phi}\right)^{-1/2}\left\{\sqrt{2\tilde{X}} - \tilde{f}(\phi) \right\} = \left(\f{2\omega + 3}{16\pi\phi}\right)^{-1/2}\tilde{\mu}~,
\end{align}
where we have used (\ref{Xeqn}),(\ref{feqn}) and Table (\ref{Einsteintable}). Finally one can check, with the above identification for $f(\phi)$, that the entropy density satisfies the predicted relation $s_J = \phi^2\tilde{s}$. 
We summarize these results in Table \ref{comp}.
\begin{table}[ht!]
    \centering
    \begin{tabular}{c c c} \toprule 
     $p_J$ & = & $\phi^2~\tilde{p}$ \\ 
    $\rho_J$ & = &  $\phi^2\tilde{\rho}$ \\ 
    $n_J$ & = &  $\phi\sqrt{\left(2\omega + 3)\phi/(16\pi\right)}\tilde{n}$ \\ 
    $\mu_J$ & = &  $\left\{\left(2\omega + 3)/(16\pi \phi\right)\right\}^{-1/2} \tilde{\mu}$ \\ 
    $h_J$ & = &  $\left\{\left(2\omega + 3)/(16\pi \phi\right)\right\}^{-1/2} \tilde{h}$ \\ 
    $s_J$ & = &  $\phi^2 \tilde{s}$ \\ 
    \bottomrule 
    \end{tabular}
    \caption{Relationship between Jordan and Einstein frame thermodynamic variables}
    \label{comp}
\end{table}

% Note that one obtains the EM tensor (\ref{ogeqn}) from the general form (\ref{emfull}) if we identify the Jordan frame variables as $u^J_{a} = \tilde{u}_a/\sqrt{\phi}$; $\rho_J = \phi^2\tilde{\rho}$ and $p_J = \phi^2\tilde{p}$ with $q=0=p_{\text{vis}} = \eta$. One can check that the normalization of velocity is preserved: $g^{ab} u_a^Ju_b^J = \phi \tilde{g}^{ab}(\tilde{u}_a\tilde{u}_b)/\phi = \tilde{g}^{ab}\tilde{u}_a\tilde{u}_b=-1$. Therefore the validity of Euler relation (\ref{Euler}) in Jordan frame implies that the terms $T_Js_J$ and $\mu_J n_J$ should be related to their counter parts in Einstein frame as $T_Js_J = \phi^2 \tilde{T} \tilde{s}$ and $\mu_J n_J = \phi^2 \tilde{\mu}\tilde{n}$. In our formulation, we choose $\tilde{T} = T_0$, which is a constant and to keep validity of the formulation in Jordan frame the temperature must be constant in this frame as well. Therefore choose $T_J = T_0 = \tilde{T}$ and then one has $s_J = \phi^2\tilde{s}$. Finally, we do not able to find a specific relation among $n_J$ and $\tilde{n}$ and similarly between $\mu_J$ and $\tilde{\mu}$. However they must be proportional to each other. The proportionality factor for $n_J$ ($\mu_J$) is either unity ($\phi^2$) or $\phi^2$ (unity) to keep $n_J$ positive.

%\subsection{Approach to GR}

Having determined all the thermodynamic variables, we can consider the GR limit of the action. This would correspond to a constant field $\psi = \psi_0$. This we will discuss in the general frame work; i.e. with respect to $\psi$ field. Einstein or Jordan frame quantities can be obtained by using the above relations.
In this case, we would have 
\begin{eqnarray}
&&p = \mc{L} = \mc{L}_0~, \,\,\,\ \rho = - p = - \mc{L}_0~, \,\,\ s = 0~;
\nonumber
\\ 
&&T = \text{constant by construction}; \,\,\,\
\mu = - f(\psi_0) = - f_0 < 0~,
\end{eqnarray}
where $\mc{L}_0$ is the value of Lagrangian density corresponding to action (\ref{5.1}) at $\psi=\psi_0$. A couple of comments are in order.
The chemical potential appears negative, which needs to be explained. As mentioned earlier, we must ensure that the function $f(\psi)$ is positive, so that the entropy density obtained is also positive (since $T > 0$ by construction). Thus, this means that the GR chemical potential is a constant, but a negative constant. Physically, the chemical potential is the energy cost of adding another particle to the system \textit{at constant entropy and volume}. For a thermodynamic system, given the total energy and the total number of particles, there is a definite number of possible arrangements of the particles which fixes the entropy through Boltzmann's definition. If we are to add a particle to this system which results in an increase of the total energy, then the number of arrangements must necessarily increase, which is to say that the entropy must increase. If we want to add a particle to the system such that the number of arrangements (and hence the entropy) is held constant, then the net energy of the system must decrease. Since the change in energy only comes from the energy of the particle added, this must mean that the energy of the particle added at constant entropy must be negative. But this is precisely how the chemical potential is defined. In this regard, recall that the chemical potential for an ideal gas in classical thermodynamics is also negative for this reason \cite{DavidTongstatmech,landau2013statistical}. The chemical potential for the ideal gas is given by (equation (2.14) of \cite{DavidTongstatmech}, or equation (45.5) of \cite{landau2013statistical})
\begin{equation}
\mu = T \text{ } \log{\left[\f{N}{V}\left(\f{2\pi\hbar^2}{mT}\right)^{\f{3}{2}}\right]} = T \text{ } \log{\left[\f{\lambda^3}{V/N}\right]}~,
\end{equation}
where 
\begin{equation}
\lambda = \sqrt{\f{2\pi\hbar^2}{mT}}~,
\end{equation}
is the thermal de-Broglie wavelength. However, the classical ideal gas law holds as long as the thermal de-Broglie wavelength $\lambda$ is much smaller than the volume per unit particle, so that we can ignore corrections from quantum mechanics; i.e.
\begin{equation}
\lambda^3 << \f{V}{N} \implies \mu < 0~. 
\end{equation}
Similarly, for a (non-interacting) Bose gas the chemical potential turns out to be negative (see Eq. (54) of \cite{landau2013statistical}). Thus, the negative chemical potential in the GR limit of the theory, is not unusual, especially when we take into account the fact that we are mapping the theory to a classical perfect fluid in thermal equilibrium at constant temperature and pressure. 

Finally, we close this section with the following comments. Note that the dictionary presented in Table \ref{dictionary} is valid only for perfect fluid case. The perfect fluid structure is usually guaranteed for a gravity-matter theory in which the matter is minimally coupled to gravity. Therefore this dictionary is useful to such a theory of gravity. However the applicability of Table \ref{dictionary}, in principle, can be generalized to any theory which {\it provides an ideal fluid EM tensor}. Therefore in summary such a dictionary is viable to any perfect fluid. In the present discussion we have used this to a general scalar-tensor theory with a non-minimally coupled scalar field, described in Jordan frame. The results are summarized in Tables \ref{Einsteintable}, \ref{jordantable} and \ref{comp}. Here we have been able to use Table \ref{dictionary} for such a theory because our identified EM tensors, both in Jordan and Einstein frames, are similar in structure to that of an ideal fluid. In that sense the outcomes, presented in Tables \ref{Einsteintable}, \ref{jordantable} and \ref{comp}, as far as our way of identifying the EM tensor for scalar fluid is concerned, are valid for any general scalar-tensor theory in which scalar field is non-minimally coupled to gravity in Jordan frame and hence they are not restricted to Brans-Dicke type.

\section{Conclusion}
We revisited the thermodynamic description of scalar fluid in ST theory of gravity using a general approach to the thermodynamics of relativistic fluids, identifying different thermodynamic and fluid parameters. Contrary to the existing works \cite{PhysRevD.103.L121501, PhysRevD.104.124031, faraoni2022scalar}, the fluid energy-momentum (EM) tensor has been identified through the description of equation of motion for metric in ST theory as thermodynamic identity as well as fluid equation. To find such EM the main emphasis has been given on the equivalent analysis of thermodynamics and fluid descriptions of gravitational dynamics on a generic null surface  both in Jordan and Einstein frames (see e.g. \cite{Bhattacharya:2020wdl,Dey:2021rke,Bhattacharya:2022mnb}). It appeared that the EM tensors in both the frames are connected by a conformal factor (see Eq. (\ref{ogeqn})) and both take ideal fluid structure. Before identifying the fluid parameters and thermodynamic quantities of these fluids, we reinvestigated the existing formulation of the same for ideal fluid. It appeared that certain alternatives can be put forwarded so that few existing short comings can improved. Importantly we found Eckart's frame may not be the best one to analysis perfect fluid.

In this paper, we have provided a possible alternate approach to the thermodynamic description of ideal scalar fluid which is valid in other than Eckart's frame. Using this the fluid and thermodynamic parameters were obtained in both the frames. Moreover, the relations among the quantities on different frames were given.  
This gives us an equivalent picture with a unique relation between thermodynamic quantities in both frames. Thus the analysis in either of the frames is sufficient to get a complete description. Importantly, we can identify a unique GR limit as the zero entropy state for the fluid i.e. the fluid entropy density vanishes in the limit of Einstein gravity. This holds in both the frames. Interestingly, in this limit the chemical potential reduces to a constant negative value, much like the chemical potential for an ideal gas in classical thermodynamics.

Finally, we mention that one should treat our investigation as one of the possibilities as far as the ideal fluid is concerned. Therefore one needs to further study to find a concrete analogy between the thermodynamic and fluid description within the perfect fluid level as well as the scalar fluid in ST theory. Hence we hope the present study is one step forward towards this and may help in future for further progress.

\vskip 3mm
{\bf Acknowledgement:}
The research of all the authors is supported by Science and Engineering Research Board (SERB), Department of Science $\&$ Technology (DST), Government of India, under the scheme Core Research Grant (File no. CRG/2020/000616). The authors also thank Valerio Faraoni for giving valuable and insightful comments on our first draft.

% \subsection{Example: Scalar Tensor Theory in Einstein frame}

% We have the action, 

% \begin{equation}
%     \mc{S}_{\psi} = \int \sqrt{-g}\text{ }d^4x\text{ }\left[\left(\f{2\omega(\psi) + 3}{16\pi}\right)X - U(\psi)\right]
% \end{equation}

% where 

% \begin{align}
%     X &= -\f{1}{2}\nabla_i\psi \nabla^i \psi \\ 
%     U(\psi) &= \f{V(\psi)}{16\pi e^{\psi}}
% \end{align}

% \begin{equation}
%     \mc{S}_{\text{EF}} = \int d^4x \text{ }\sqrt{-g}\left[\f{\tilde{R}}{16 \pi} - \left(\f{2\omega(\psi) + 3}{16\pi}\right)\f{1}{2}\tilde{g}^{ab}\nabla_a\tilde{\psi}\nabla_b\tilde{\psi} - U(\tilde{\psi})\right]
% \end{equation}

% with \[U(\tilde{\psi}) = \f{V(\psi)}{16\pi e^{\psi}}\]where the tilde variables indicate Einstein frame quantities. We take this so as to obtain the $T_{ab}$ mentioned in \ref{ogeqn97}. From this, we immediately identify the required quantities as, 

% \begin{enumerate}
%     \item $p = \mc{L}$
%     \item $\rho = 2X\left(2\omega(\psi) + 3\right)/16\pi - \mc{L} $ 
    
% \end{enumerate}

\appendix 
\begin{center}
\section*{Appendices}
\end{center}

\section{Derivation of Eq. (\ref{5.14})}\label{AppB}
Euler relation (\ref{Euler}). This yields
\begin{equation}
\nabla_a\mu = \nabla_a\left(\f{\rho + p}{n}\right) - T_0 \nabla_a\left(\f{s}{n}\right) = \nabla_a\left(\f{\rho + p}{n}\right) - T_0\left[\f{\nabla_a\rho}{nT_0} - \f{p + \rho}{n^2T_0}\nabla_an\right] = \f{\nabla_a p}{n}~,
 \label{5.11}
\end{equation}
where in the second equality we have used the following identity
\begin{equation}
n\nabla_a\left(\f{s}{n}\right) = \f{\nabla_a\rho}{T_0} - \f{p+\rho}{nT_0} \nabla_a n~,
\label{B.2}
\end{equation}
which is derived as follows. From the first law of thermodynamics (per unit volume) one finds.
\begin{align}
&\nabla_a\rho = T_0\nabla_as + \mu \nabla_an = T_0 \nabla_a\left(n \f{s}{n}\right) + \mu \nabla_an 
\nonumber 
\\
&= n T \nabla_a\left(\f{s}{n}\right) + \left(\mu + T \f{s}{n}\right)\nabla_an~.
\end{align}
Next use of Euler's relation (\ref{Euler}) yields (\ref{B.2}).
Now using (\ref{dict1}) and (\ref{dict2}) we find $\nabla_ap = \nabla_a\mc{L} = \mc{L}_{\psi}\nabla_a\psi + \mc{L}_X\nabla_aX$ and so one has
\begin{align}
\f{\nabla_ap}{n} &= \f{1}{n}\left(\sqrt{2X}\mc{L}_X\text{ }u_a + \mc{L}_X\nabla_aX\right)
\nonumber 
\\ 
&= \left(\f{\mc{L}_{\psi}}{\mc{L}_X}\right)u_a + \nabla_a\left(\sqrt{2X}\right)~.
\label{5.13}
\end{align}
Then combining (\ref{5.11}) and (\ref{5.13}) we can take the spatial projection to obtain (\ref{5.14})

%\newpage 
\bibliographystyle{IEEEtran}
\bibliography{bibl.bib}

\begin{thebibliography}{10}
\providecommand{\url}[1]{#1}
\csname url@samestyle\endcsname
\providecommand{\newblock}{\relax}
\providecommand{\bibinfo}[2]{#2}
\providecommand{\BIBentrySTDinterwordspacing}{\spaceskip=0pt\relax}
\providecommand{\BIBentryALTinterwordstretchfactor}{4}
\providecommand{\BIBentryALTinterwordspacing}{\spaceskip=\fontdimen2\font plus
\BIBentryALTinterwordstretchfactor\fontdimen3\font minus
  \fontdimen4\font\relax}
\providecommand{\BIBforeignlanguage}[2]{{%
\expandafter\ifx\csname l@#1\endcsname\relax
\typeout{** WARNING: IEEEtran.bst: No hyphenation pattern has been}%
\typeout{** loaded for the language `#1'. Using the pattern for}%
\typeout{** the default language instead.}%
\else
\language=\csname l@#1\endcsname
\fi
#2}}
\providecommand{\BIBdecl}{\relax}
\BIBdecl

\bibitem{Padmanabhan:2003gd}
T.~Padmanabhan, ``{Gravity and the thermodynamics of horizons},'' \emph{Phys.
  Rept.}, vol. 406, pp. 49--125, 2005.

\bibitem{Padmanabhan:2014jta}
------, ``{Emergent Gravity Paradigm: Recent Progress},'' \emph{Mod. Phys.
  Lett. A}, vol.~30, no. 03n04, p. 1540007, 2015.

\bibitem{Kolekar:2011gw}
S.~Kolekar and T.~Padmanabhan, ``{Action Principle for the Fluid-Gravity
  Correspondence and Emergent Gravity},'' \emph{Phys. Rev. D}, vol.~85, p.
  024004, 2012.

\bibitem{Padmanabhan:2013xyr}
T.~Padmanabhan and D.~Kothawala, ``{Lanczos-Lovelock models of gravity},''
  \emph{Phys. Rept.}, vol. 531, pp. 115--171, 2013.

\bibitem{Jacobson:1995ab}
T.~Jacobson, ``{Thermodynamics of space-time: The Einstein equation of
  state},'' \emph{Phys. Rev. Lett.}, vol.~75, pp. 1260--1263, 1995.

\bibitem{Eling:2006aw}
C.~Eling, R.~Guedens, and T.~Jacobson, ``{Non-equilibrium thermodynamics of
  spacetime},'' \emph{Phys. Rev. Lett.}, vol.~96, p. 121301, 2006.

\bibitem{Callan:1985ia}
C.~G. Callan, Jr., E.~J. Martinec, M.~J. Perry, and D.~Friedan, ``{Strings in
  Background Fields},'' \emph{Nucl. Phys. B}, vol. 262, pp. 593--609, 1985.

\bibitem{Esposito-Farese:2003jfi}
G.~Esposito-Farese, ``{Scalar tensor theories and cosmology and tests of a
  quintessence Gauss-Bonnet coupling},'' in \emph{{38th Rencontres de Moriond
  on Gravitational Waves and Experimental Gravity}}, 6 2003.

\bibitem{Elizalde:2004mq}
E.~Elizalde, S.~Nojiri, and S.~D. Odintsov, ``{Late-time cosmology in (phantom)
  scalar-tensor theory: Dark energy and the cosmic speed-up},'' \emph{Phys.
  Rev. D}, vol.~70, p. 043539, 2004.

\bibitem{Saridakis:2016ahq}
E.~N. Saridakis and M.~Tsoukalas, ``{Cosmology in new gravitational
  scalar-tensor theories},'' \emph{Phys. Rev. D}, vol.~93, no.~12, p. 124032,
  2016.

\bibitem{Crisostomi:2016czh}
M.~Crisostomi, K.~Koyama, and G.~Tasinato, ``{Extended Scalar-Tensor Theories
  of Gravity},'' \emph{JCAP}, vol.~04, p. 044, 2016.

\bibitem{Langlois:2017dyl}
D.~Langlois, R.~Saito, D.~Yamauchi, and K.~Noui, ``{Scalar-tensor theories and
  modified gravity in the wake of GW170817},'' \emph{Phys. Rev. D}, vol.~97,
  no.~6, p. 061501, 2018.

\bibitem{Faraoni:1999hp}
V.~Faraoni and E.~Gunzig, ``{Einstein frame or Jordan frame?}'' \emph{Int. J.
  Theor. Phys.}, vol.~38, pp. 217--225, 1999.

\bibitem{Bhattacharya:2017pqc}
K.~Bhattacharya and B.~R. Majhi, ``{Fresh look at the scalar-tensor theory of
  gravity in Jordan and Einstein frames from undiscussed standpoints},''
  \emph{Phys. Rev. D}, vol.~95, no.~6, p. 064026, 2017.

\bibitem{Bhattacharya:2018xlq}
K.~Bhattacharya, A.~Das, and B.~R. Majhi, ``{Noether and Abbott-Deser-Tekin
  conserved quantities in scalar-tensor theory of gravity both in Jordan and
  Einstein frames},'' \emph{Phys. Rev. D}, vol.~97, no.~12, p. 124013, 2018.

\bibitem{Bhattacharya:2020jgk}
K.~Bhattacharya, ``{Thermodynamic aspects and phase transition of black
  holes},'' Ph.D. dissertation, Gauhati U., 2020.

\bibitem{Bhattacharya:2022mnb}
K.~Bhattacharya and B.~R. Majhi, ``{Scalar\textendash{}tensor gravity from
  thermodynamic and fluid-gravity perspective},'' \emph{Gen. Rel. Grav.},
  vol.~54, no.~9, p. 112, 2022.

\bibitem{Steinwachs:2011zs}
C.~F. Steinwachs and A.~Y. Kamenshchik, ``{One-loop divergences for gravity
  non-minimally coupled to a multiplet of scalar fields: calculation in the
  Jordan frame. I. The main results},'' \emph{Phys. Rev. D}, vol.~84, p.
  024026, 2011.

\bibitem{Kamenshchik:2014waa}
A.~Y. Kamenshchik and C.~F. Steinwachs, ``{Question of quantum equivalence
  between Jordan frame and Einstein frame},'' \emph{Phys. Rev. D}, vol.~91,
  no.~8, p. 084033, 2015.

\bibitem{Banerjee:2016lco}
N.~Banerjee and B.~Majumder, ``{A question mark on the equivalence of Einstein
  and Jordan frames},'' \emph{Phys. Lett. B}, vol. 754, pp. 129--134, 2016.

\bibitem{Pandey:2016unk}
S.~Pandey and N.~Banerjee, ``{Equivalence of Jordan and Einstein frames at the
  quantum level},'' \emph{Eur. Phys. J. Plus}, vol. 132, no.~3, p. 107, 2017.

\bibitem{Ruf:2017xon}
M.~S. Ruf and C.~F. Steinwachs, ``{Quantum equivalence of $f(R)$ gravity and
  scalar-tensor theories},'' \emph{Phys. Rev. D}, vol.~97, no.~4, p. 044050,
  2018.

\bibitem{Bhattacharya:2020wdl}
K.~Bhattacharya, B.~R. Majhi, and D.~Singleton, ``{Fluid-gravity correspondence
  in the scalar-tensor theory of gravity: (in)equivalence of Einstein and
  Jordan frames},'' \emph{JHEP}, vol.~07, p. 018, 2020.

\bibitem{Dey:2021rke}
S.~Dey, K.~Bhattacharya, and B.~R. Majhi, ``{Thermodynamic structure of a
  generic null surface and the zeroth law in scalar-tensor theory},''
  \emph{Phys. Rev. D}, vol. 104, no.~12, p. 124038, 2021.

\bibitem{Faraoni:1998qx}
V.~Faraoni, E.~Gunzig, and P.~Nardone, ``{Conformal transformations in
  classical gravitational theories and in cosmology},'' \emph{Fund. Cosmic
  Phys.}, vol.~20, p. 121, 1999.

\bibitem{Faraoni:2010yi}
V.~Faraoni, ``{Black hole entropy in scalar-tensor and f(R) gravity: An
  Overview},'' \emph{Entropy}, vol.~12, p. 1246, 2010.

\bibitem{PhysRevD.103.L121501}
\BIBentryALTinterwordspacing
V.~Faraoni and A.~Giusti, ``Thermodynamics of scalar-tensor gravity,''
  \emph{Phys. Rev. D}, vol. 103, p. L121501, Jun 2021. [Online]. Available:
  \url{https://link.aps.org/doi/10.1103/PhysRevD.103.L121501}
\BIBentrySTDinterwordspacing

\bibitem{PhysRevD.104.124031}
\BIBentryALTinterwordspacing
V.~Faraoni, A.~Giusti, and A.~Mentrelli, ``New approach to the thermodynamics
  of scalar-tensor gravity,'' \emph{Phys. Rev. D}, vol. 104, p. 124031, Dec
  2021. [Online]. Available:
  \url{https://link.aps.org/doi/10.1103/PhysRevD.104.124031}
\BIBentrySTDinterwordspacing

\bibitem{faraoni2022scalar}
V.~Faraoni, S.~Giardino, A.~Giusti, and R.~Vanderwee, ``Scalar field as a
  perfect fluid: thermodynamics of minimally coupled scalars and einstein frame
  scalar-tensor gravity,'' \emph{arXiv preprint arXiv:2208.04051}, 2022.

\bibitem{PhysRev.58.919}
\BIBentryALTinterwordspacing
C.~Eckart, ``The thermodynamics of irreversible processes. iii. relativistic
  theory of the simple fluid,'' \emph{Phys. Rev.}, vol.~58, pp. 919--924, Nov
  1940. [Online]. Available:
  \url{https://link.aps.org/doi/10.1103/PhysRev.58.919}
\BIBentrySTDinterwordspacing

\bibitem{Warner}
W.~Israel, ``{Gravity and the thermodynamics of horizons},'' \emph{Annals of
  Phys.}, vol. 100, pp. 310--331, 1976.

\bibitem{israel1979transient}
W.~Israel and J.~M. Stewart, ``Transient relativistic thermodynamics and
  kinetic theory,'' \emph{Annals of Physics}, vol. 118, no.~2, pp. 341--372,
  1979.

\bibitem{Koga:1998un}
J.-i. Koga and K.-i. Maeda, ``{Equivalence of black hole thermodynamics between
  a generalized theory of gravity and the Einstein theory},'' \emph{Phys. Rev.
  D}, vol.~58, p. 064020, 1998.

\bibitem{Chakraborty:2015aja}
S.~Chakraborty, K.~Parattu, and T.~Padmanabhan, ``{Gravitational field
  equations near an arbitrary null surface expressed as a thermodynamic
  identity},'' \emph{JHEP}, vol.~10, p. 097, 2015.

\bibitem{Dey:2020tkj}
S.~Dey and B.~R. Majhi, ``{Covariant approach to the thermodynamic structure of
  a generic null surface},'' \emph{Phys. Rev. D}, vol. 102, no.~12, p. 124044,
  2020.

\bibitem{Price:1986yy}
R.~H. Price and K.~S. Thorne, ``{Membrane Viewpoint on Black Holes: Properties
  and Evolution of the Stretched Horizon},'' \emph{Phys. Rev. D}, vol.~33, pp.
  915--941, 1986.

\bibitem{Parikh:1997ma}
M.~Parikh and F.~Wilczek, ``{An Action for black hole membranes},'' \emph{Phys.
  Rev. D}, vol.~58, p. 064011, 1998.

\bibitem{Gourgoulhon:2005ng}
E.~Gourgoulhon and J.~L. Jaramillo, ``{A 3+1 perspective on null hypersurfaces
  and isolated horizons},'' \emph{Phys. Rept.}, vol. 423, pp. 159--294, 2006.

\bibitem{Padmanabhan:2010rp}
T.~Padmanabhan, ``{Entropy density of spacetime and the Navier-Stokes fluid
  dynamics of null surfaces},'' \emph{Phys. Rev. D}, vol.~83, p. 044048, 2011.

\bibitem{andersson2007relativistic}
N.~Andersson and G.~L. Comer, ``Relativistic fluid dynamics: Physics for many
  different scales,'' \emph{Living Reviews in Relativity}, vol.~10, no.~1, pp.
  1--83, 2007.

\bibitem{pimentel1989energy}
L.~O. Pimentel, ``Energy-momentum tensor in the general scalar-tensor theory,''
  \emph{Classical and Quantum Gravity}, vol.~6, no.~12, p. L263, 1989.

\bibitem{PhysRevD.98.084019}
\BIBentryALTinterwordspacing
V.~Faraoni and J.~C\^ot\'e, ``Imperfect fluid description of modified
  gravities,'' \emph{Phys. Rev. D}, vol.~98, p. 084019, Oct 2018. [Online].
  Available: \url{https://link.aps.org/doi/10.1103/PhysRevD.98.084019}
\BIBentrySTDinterwordspacing

\bibitem{piattella2014note}
O.~F. Piattella, J.~C. Fabris, and N.~Bili{\'c}, ``Note on the thermodynamics
  and the speed of sound of a scalar field,'' \emph{Classical and Quantum
  Gravity}, vol.~31, no.~5, p. 055006, 2014.

\bibitem{kremer2014diffusion}
G.~M. Kremer, ``Diffusion of relativistic gas mixtures in gravitational
  fields,'' \emph{Physica A: Statistical Mechanics and its Applications}, vol.
  393, pp. 76--85, 2014.

\bibitem{DavidTongstatmech}
\BIBentryALTinterwordspacing
D.~Tong, ``Lectures on general relativity.'' [Online]. Available:
  \url{http://www.damtp.cam.ac.uk/user/tong/statphys.html}
\BIBentrySTDinterwordspacing

\bibitem{landau2013statistical}
L.~D. Landau and E.~M. Lifshitz, \emph{Statistical Physics: Volume 5}.\hskip
  1em plus 0.5em minus 0.4em\relax Elsevier, 2013, vol.~5.

\end{thebibliography}
\end{document}